\documentclass[showpacs,twocolumn,floatfix]{revtex4}
\usepackage{graphicx}
\usepackage{amssymb}
\usepackage{mathpazo}
\usepackage{epstopdf}
\usepackage{float}
\usepackage{amsmath}

\newcommand{\ignore}[1]{}

\newcommand{\be}{\begin{equation}}
\newcommand{\ee}{\end{equation}}
\def\ba#1\ea{\begin{align}#1\end{align}}
\newcommand{\bit}{\begin{itemize}}
\newcommand{\eit}{\end{itemize}}

\newcommand{\nn}{\nonumber}

\renewcommand{\a}{\alpha}

\def\slashb#1{\setbox0=\hbox{$#1$}#1\hskip-\wd0\dimen0=5pt\advance
        \dimen0 by-\ht0\advance\dimen0 by\dp0\lower0.5\dimen0\hbox
          to\wd0{\hss\sl/\/\hss}}

\input{epsf}
\begin{document}

\title{Optimizing the Determination of \\
the Neutrino Mixing Angle $\theta_{13}$ from Reactor Data }
\author{Amir N. Khan$^{*}$, Douglas W. McKay$^{\dagger}$, and John P. Ralston%
$^{\dagger}$}
\affiliation{$^{*}$Department of Physics, COMSATS IIT, Park Road, 44000, Islamabad,
Pakistan \\
$^{\dagger}$Department of Physics \& Astronomy, the University of Kansas,
Lawrence, KS 66045 USA}

\begin{abstract}
The technical breakthroughs of multiple detectors developed by Daya Bay and
RENO collaborations have gotten great attention. Yet the optimal
determination of neutrino mixing parameters from reactor data depends on the
statistical method and demands equal attention. We find that a
straightforward method using a minimal parameters will generally outperform
a multi-parameter method by delivering more reliable values with sharper
resolution. We review standard confidence levels and statistical penalties
for models using extra parameters, and apply those rules to our analysis. We
find that the methods used in recent work of the Daya Bay and RENO
collaborations have several undesirable properties. The existing work also
uses non-standard measures of significance which we are unable to explain. A
central element of the current methods consists of variationally fitting
many more parameters than data points. As a result the experimental
resolution of $\sin ^{2}(2\theta _{13})$ is degraded. The results also
become extremely sensitive to certain model parameters that can be adjusted
arbitrarily. The number of parameters to include in evaluating significance
is an important issue that has generally been overlooked. The measures of
significance applied previously would be consistent if and only if all
parameters but one were considered to have no physical relevance for the
experiment's hypothesis test. Simpler, more transparent methods can improve
the determination of the mixing angle $\theta _{13}$ from reactor data, and
exploit the advantages from superb hardware technique of the experiments. We
anticipate that future experimental analysis will fully exploit those
advantages.
\end{abstract}

\maketitle


\section{A Technological Breakthrough}

\label{sec:intro}

It goes without saying that experiments with great technical accomplishment
should be evaluated with data analysis of equal or better quality. During
the past year or so, the achievement of constructing multiple, nearly
identical neutrino detectors by the Daya Bay \cite{db1,db2} and RENO \cite%
{reno} collaborations has been rightly praised as a technological
breakthrough. Beyond increasing data rates, the prime function of the new
technology is to reduce systematic errors. Systematic errors previously
dominated neutrino oscillation experiments with nuclear reactor sources for
many years. Yet by a curious gap in the current literature, the data
analysis published to quantify the neutrino mixing angle \cite{pont,mns} $%
\theta _{13}$ is far from optimal. Applying more effective methods to the
analysis can yield higher resolution of neutrino physics parameters than
currently available. Despite lacking complete access to the full
information, we can make a case for producing better determination of $\sin
^{2}(2\theta _{13})$ and its uncertainties than the experimental reports. We
are naturally surprised by this fact. It is primarily due to unrecognized
faults in the inefficient methods used before.

The experimental uncertainties on $\theta _{13}$ have been the center of
attention for years after the CHOOZ null results of 1999 and 2003 \cite%
{chooz}. Uncertainties remained the focus after the upgraded Double Chooz 
\cite{dchooz1} report of new results just before Daya Bay's and RENO's
results, their improved, 2.9$\sigma $ result a few months later \cite%
{dchooz2}, and a recent result consistent with all previous measurements,
but using the delayed neutron capture from hydrogen for the first time \cite%
{dchooz3}. What has gone largely unnoticed is that the statistical method
used in these papers diverged significantly from most previous work, cannot
directly be compared, and shows signals of being problematic.

There usually exists more than one ``correct forms'' of data analysis. Most
physicists agree one should not be overly concerned with any method,
provided the assumptions are reported, that the method is robust under small
perturbations, and that the results are reproducible. We will present such a
method analyzing the Daya Bay data. The method includes \textit{stating a
specific hypothesis}, which may appear quaint, but if neglected leads to no
hypothesis to test. The method uses few rather than many parameters, and we
report everything needed to reproduce our calculations. Remarkably, the
current experimental literature on $\theta_{13}$ is not definite on any
hypothesis, is not reproducible, and its approach does not appear robust
under small model parameters that can be freely adjusted.

Once the ground rules are defined, quantifying confidence levels with
goodness of fit statistics becomes meaningful. Without ground rules and
reproducibility the dependence of a statistic on a parameter has little
objective meaning. The tendency to name all statistics ``$\chi^{2}$''
regardless of their actual definition does not make them all equivalent.
When the meaning and values of parameters are omitted from discussion, it is
impossible to know whether or not they are ``nuisance parameters.'' A 
\textit{textbook} nuisance parameter is one whose value is completely
irrelevant to the hypothesis, but which must be accounted for in the
analysis of the parameters that are relevant. Meanwhile there are few if any
textbook nuisance parameters in experimental physics. Every parameter has a
physical meaning. If a data fit finds a parameter far from expectations, it
indicates something is wrong, whether or not the nuisance is annoying. Such
a nuisance parameter can invalidate the entire study, depending on its
value. Unless one finds a reason otherwise, it is unavoidable that both the
``uninteresting'' and the ``interesting'' parameters contribute to the
actual hypothesis and its uncertainties.

The upshot is that using extra parameters will carry a statistical penalty
if one cares about them. Extra parameters should not be used if one does not
care about them. Having it both ways is a impossible for us to defend. After
the first version of this paper \cite{amirv1}, we undertook a literature
search to review the history. It turns out that the references of the Daya
Bay experimental proposal \cite{DBproposal} actually employed a more
conservative determination of confidence regions, consistent with ours and
contradicting the method Daya Bay used when the data appeared. This is
discussed further in Section \ref{sec:regions}.

Section \ref{sec:straight} begins with a simple straightforward procedure
with a clearly stated hypothesis including a list of parameters central to
the hypothesis. We will discover an opportunity to retrospectively
re-classify a parameter as a nuisance after it was fit and found consistent
with expectations. Since that step would abolish the original test
conditions, we cannot find a way to justify it. The hypothetical case that
our fit stands on a better footing than Daya Bay' hinges on the fact we
account for all our parameters and pay the statistical penalty up front:
plus our calculations are reproducible. Section \ref{sec:chipull} sets up an
illustration of the \textit{method of $\chi ^{2}$ with pull }\cite{fogli1,
sksolar, strumia, degouvea, huber03} that has become the exclusive tool of
analysis by the experimental collaborations cited. The method uses many
parameters of physical importance, and also turns out to be remarkably
sensitive to fine details of tuning external parameters. We explore the
method while stating our assumptions, sticking to them, and also provide all
the information to reproduce our calculations. This fills a gap in the
neutrino literature where the procedures of assigning errors have not been
spelled out for the users of the data. When we compare our results following
standard procedures with the number of data points and (very large) number
of parameters used by Daya Bay and RENO there is a unexplained discrepancy.
We cannot explain what hypothesis those experimental groups are assuming,
nor find it stated anywhere. The Section also explains how, paradoxically, a
definite \textit{insensitivity} of $\chi ^{2}$ defined in that approach is
not the virtue it appears. Excessive parameters tends to degrade the
determination of the physical objective, $\sin ^{2}(2\theta _{13})$. Due to
this situation, there is enough leeway in the current determination of $%
\theta _{13}$ to make two logical but contradictory arguments. It is
possible to find the uncertainty of $\theta _{13}$ has been greatly \textit{%
underestimated}, and it is also possible to find the uncertainty has been
significantly \textit{overestimated}. Though we are obviously not in a
position to resolve the alternatives, we find it fascinating to understand
the issues and develop means to assess the situation. That leads to our main
conclusion (Section \ref{sec:discuss}) that simpler methods are preferred,
both for scientific and mathematical reasons. An Appendix gives details on
how we extracted data from the publications.

\section{Errors Depend on the Procedure}

\label{sec:straight}

\subsubsection{A Straightforward, Simple Approach}

For our first example we present a straightforward, simple model. We use the
Daya Bay ($DB$) data of Ref. \cite{db1} to illustrate the concepts.

\begin{figure}[htbp]
\begin{center}
\includegraphics[width=3in]{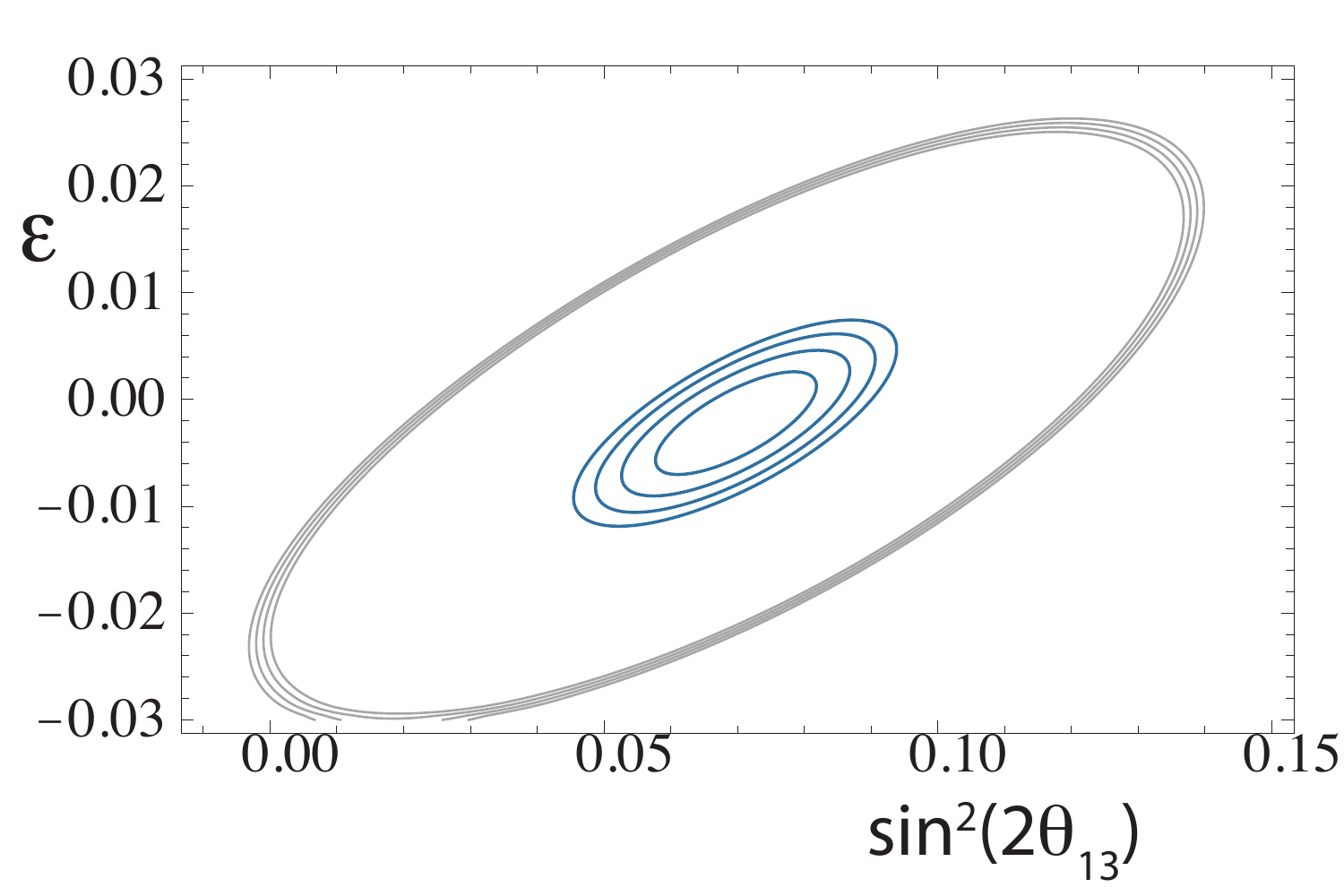}
\end{center}
\caption{ Contours of constant $\chi^{2}$ of the simple 2-parameter
fit, Eq. \ref{simple}. Inner contours (blue online) are the minimum
value of 5.66 plus 1, 2, 3, 4 units. Outer contours are also separated by 1
unit and begin at $\Delta \chi^{2}=37.6$ units, the value for the
null hypothesis $sin^{2}(2 \theta_{13})=0$. The significance of $%
\Delta \chi^{2}=32$ evaluated using a 2-parameter fit is the
equivalent of 5.4 $\sigma$. }
\label{fig:SimpleFit.eps}
\end{figure}

The object of the exercise is to determine $\sin^{2}(2 \theta_{13})
=s_{213}^{2}$ using a model $M_{d\, Model} $ for the $d$th detector. The
model assumes a certain reactor flux and detector efficiency, which have a
parameter $\epsilon$ describing its relative uncertainty. We cannot avoid $%
\epsilon$ and need to determine it self-consistently. Our null hypothesis is
that $\epsilon$ is of order $1\% \pm 1\%$, and that $s_{213}^{2}=0.$ The
point of fitting the data will be to find whether the null model can be
ruled out, and compute confidence levels on the two parameters fit relative
to the null model used.

Let $M_d$ be the total number of events seen in the $d$-th detector. We
define a statistic \ba \chi^{2} & = \sum_{d=1}^{dmax}\frac{(M_{d}-M_{d
\,Model})^{2}}{ M_{d}}; \nonumber \\
M_{d\, Model} &=(1-s_{213}^{2} s_{d}^{2}) M_{d\, 0}(1+\epsilon) \nonumber 
\\ &=P_{d}M_{d\,0}(1+\epsilon). \label{simple} \ea 

Here $M_{d0}$ is the number of events expected with no oscillation, $%
s_{d}^{2}=<\sin^{2}(\Delta m_{13}^{2}L_{d}/E_{\nu} ) >$ is averaged over the
energy flux at flux-weighted reactor-detector separation $L_{d}$. The
denominator $(\sqrt{M_{d}})^{2}$ is the conventional variance from Poisson
statistics.

Consider a one-detector experiment like Double Chooz \cite{dchooz1,dchooz2}.
Then $(1-s_{213}^{2} s_{1}^{2}) M_{1\, 0}(1+\epsilon)$ in Eq. \ref{simple}
is a single number that cannot determine two variables $s_{213}^{2}$ and $%
\epsilon$. Due to that degeneracy the ignorance in reactor flux and detector
response directly translates into systematic error in $s_{213}^{2}$, and
neither can be determined unambiguously.

Consider an experiment like RENO \cite{reno} with $2$ detectors at $2$
well-chosen baselines. With 2 data points the degeneracy is removed, but
parameters are just barely determined, not over-determined. To a good
approximation Daya Bay also has 2 flux-weighted baselines, as one can see
from their Figure 4, reproduced here as Fig. 2. There is a near set at $%
(0.467,0.474,0.578)$ km and a far set at$1.647$ km (the separations of the
far set are for visualization). The existence of $3$ near and $3$ far
detectors effectively triples the amount of data. Due to this situation we
do not anticipate a fit to two parameters will be over-determined.

{\small 
\begin{table}[htbp]
\begin{center}
{\small 
\begin{tabular}{ccccccc}
$Detector $ & 1 & 2 & 3 & 4 & 5 & 6 \\ \hline
$M_{d0} $ & 28647 & 29096 & 22335 & 3567 & 3573 & 3536 \\ 
$P_{d}$ & 0.991 & 0.977 & 0.987 & 0.941 & 0.929 & 0.913 \\ 
$M_{d } $ & 28389 & 28427 & 22045 & 3356.5 & 3319 & 3228 \\ 
$L_{eff \, d}$ & 0.474 & 0.467 & 0.578 & 1.647 & 1.647 & 1.647 \\ 
&  &  &  &  &  & 
\end{tabular}
}
\end{center}
\par
{\small \label{tab:tableI}  }
\caption{Data used in fits. Units of $L_{eff \, d}$ are km.}
\end{table}
}

We fit the simple model with the data shown in Table I, derived in the
Appendix. The fit gives $\chi ^{2}=5.66$ at $\epsilon
=-0.002,\,s_{213}^{2}=0.070\pm 0.018$. A difference of nearly $32$ units of $%
\chi ^{2}$ separate $\chi ^{2}(s_{213}^{2}=0),\epsilon =0.01$ from the best
fit value. We emphasize that both $\epsilon $ and $s_{213}^{2}$ are
meaningful, so that the standard evaluation of significance of
\textquotedblleft detection\textquotedblright\ uses $\chi _{2}^{2}$. We
review the reasoning behind this next.

\subsubsection{Defining Measures of Significance}

\label{sec:regions}

When data fitting a model comes from a Gaussian distribution, or more
generally any distribution with a suitably isolated ``bump,'' then the
statistic $z=\chi^{2}$ is predicted to be distributed by the $\chi_{\nu}(z)$
distribution: \ba \chi_{\nu}(z) = \frac{2^{-\nu /2}}{\Gamma(\nu /2)}
e^{-z/2} z^{\frac{\nu }{2}-1}. \nonumber \ea 

The estimated number of degrees of freedom $\nu=N-p$ when there are $N$
terms in $\chi^{2}$ and $p$ parameters. We have $N=6$ and $p=2$ hence $\nu=4$%
, as far as the best-fit is concerned. But rather than focusing on value of $%
\chi^{2}$, we are concerned with the \textit{difference} $\Delta \chi^{2}$
between the values of two hypotheses. When two models are nested, meaning
one is smoothly immersed in the other by varying $q$ parameters, Wilks'
Theorem predicts $\Delta \chi^{2}$ is distributed by $\chi_{q}^{2}$. The
theorem is more general than assuming a Gaussian distribution, but that is
not our point. For now, we are emphasizing \textit{the decision to use $%
\nu=2 $ for the specific 2-parameter question of ``detection.''} It is
supported by a theorem, and we confirmed it by simulations, yet assessment
with $\nu=2$ is a \textit{decision based on definite assumptions} that we
have listed.

The outcome then rejects the null hypothesis $s_{213}^{2}=0$ by the Gaussian
equivalent of 5.4$\sigma$. The result is quite close to the significance of
5.2$\sigma$ reported by $DB$ for the same data set. However we will soon see
this is a coincidence because $DB$'s criteria of significance are much
different from those we illustrate here.

\begin{figure}[htbp]
\begin{center}
\includegraphics[width=3in]{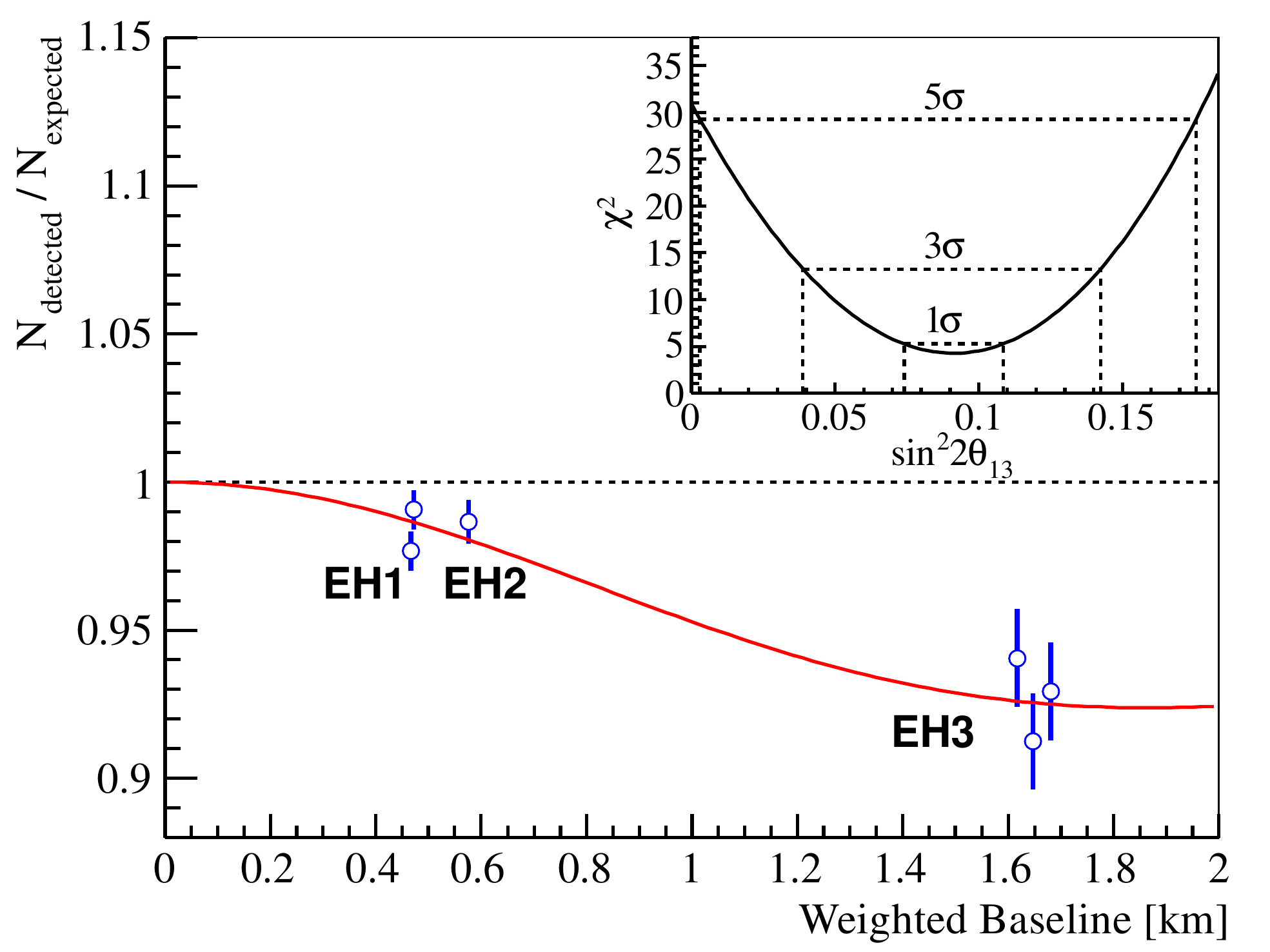}
\end{center}
\caption{{\small Daya Bay's Figure 4, showing the survival
probability at best fit. The 1$\sigma$, 3$\sigma$, and 5 $%
\sigma$ confidence intervals of the inset indicate significance is
evaluated using the standard one-parameter criterion $\chi_{1}^{2}$.
Digitizing the red curve gives the energy and efficiency-averaged detector
response. }}
\label{fig:DB_oscillation.eps}
\end{figure}

\textit{Parameter uncertainties} are quite a different thing from testing
hypotheses. In Section \ref{sec:intro} we mentioned the practice of citing
parameter uncertainties using $\Delta \chi ^{2}=1$. When a fit uses only one
(1) parameter the 68\% confidence interval coming from $\chi _{1}^{2}$ (note
subscript \textquotedblleft 1')' is indeed the range where $\Delta \chi
^{2}=1$. Our simple fit uses 2 parameters which jointly need to be
monitored. Then it is standard practice \cite{pdgstats,recipes} to evaluate
significance \textit{of the two-dimensional parameter region} using the $%
\chi _{2}^{2}$ distribution. (Note subscript \textquotedblleft
2\textquotedblright ). This statistical penalty takes into account the extra
freedom for either parameter to \textquotedblleft float\textquotedblright\
while the other is varied. The error ellipse from two parameters requires $%
\Delta \chi _{2}^{2}=(2.3,\,6.17,\,11.8)$ to generate $(1\sigma ,\,2\sigma
,\,3\sigma $) confidence levels. Following standard practice we effectively
used the contours of $\chi ^{2}=\chi _{min}^{2}+2.3=7.95$ in reporting the
uncertainty in $s_{213}^{2}=0.070\pm 0.018$, which can be checked with
Figure \ref{fig:SimpleFit.eps}. Notice that this contour crosses close to
the prior value of $\epsilon \sim 0.01$. The self-consistency of $\epsilon $
gives confidence that the value of $\,\,\,s_{213}^{2}$ is reliable.

\begin{figure}[tbp]
\begin{center}
\includegraphics[width=3in]{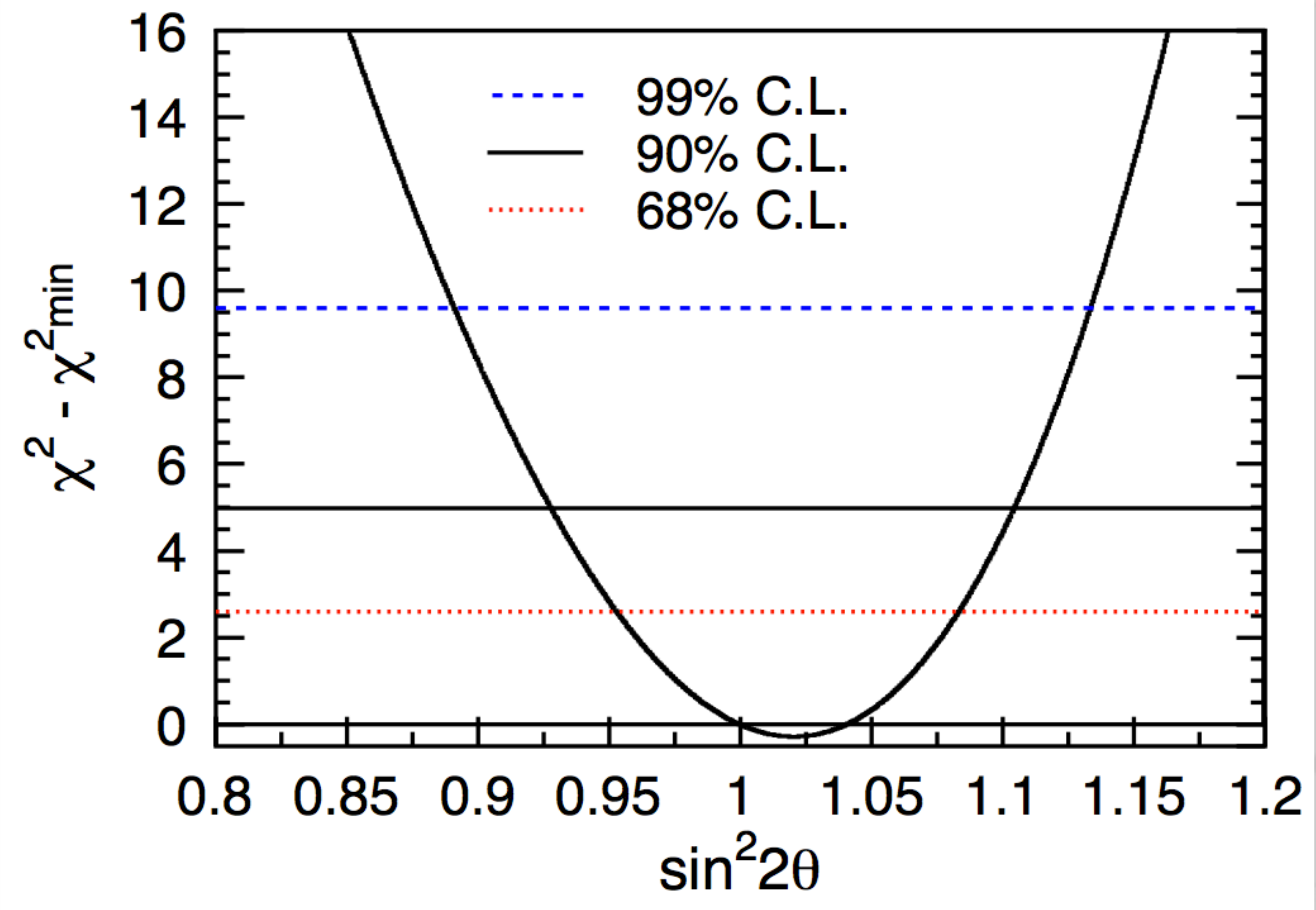}
\end{center}
\caption{{\small A panel from Figure 39 of Ashie et al \cite%
{ashie} showing confidence levels used for a 2-parameter fit. Compare the 1$%
\sigma$ (68\%) and 3$\sigma$ (99.5\%) levels of the inset of
Figure \ref{fig:DB_oscillation.eps}. }}
\label{fig:AshieFig39.pdf}
\end{figure}

Our reason for dwelling on $\epsilon $ does not come from intimate knowledge
of the hardware. We care about $\epsilon $ because if the central value fit
had been 0.03, for example, we'd distrust the value of $s_{213}$ and its
error bars. In no way could we call $\epsilon $ a nuisance parameter.
Support for our procedure comes from Fukuda \textit{et al. }\cite{fukuda}
and Ashie \textit{et al. }\cite{ashie}, which are concerned with jointly
fitting $\nu _{\mu }\rightarrow \nu _{\tau }$ oscillations with two
parameters $\Delta m^{2}$ and $\sin ^{2}(2\theta )$ . Both papers make an
explicit statement that the 2-parameter procedure uses different criteria
than a 1-parameter procedure. Both papers were cited for the statistical
method by the Daya Bay experimental proposal. These papers cite $\Delta \chi
^{2}=$ 2.6, 4.6, 9.6 for 68\%, (1$\sigma $) 90\% and 99\% ($\sim 3\,\sigma $%
) confidence regions \footnote{%
The exact numbers are quite sensitive. To 5-digit accuracy we find 0.72747,
0.89974, 0.99177 significance.}. Figure \ref{fig:AshieFig39.pdf} is taken
from Figure 39 of Ref. \cite{ashie} and shows how the $n=\sigma $ lines
differ from the inset of Figure \ref{fig:DB_oscillation.eps}.


Given that $\epsilon =-0.002$ is very small, we might have retrospectively
set it to zero, and re-fit the data to one parameter. That sounds like
cheating. However if one had been highly confident that $\epsilon =0$, or
any other number with negligible uncertainty, it is legitimate to state that
information as a definite hypothesis. Under that new hypothesis a
one-parameter fit is made with $\epsilon $ fixed and varying $s_{123}^{2}$.
The value of having high quality advance information and a one-parameter
hypothesis is that $\Delta \chi ^{2}$ is distributed by $\chi _{1}^{2}$.
Using $\chi _{1}^{2}$ for assessment only needs $\Delta \chi
^{2}=1,\,4,\,9...$ for the equivalent $1\sigma ,\,2\sigma ,\,3\sigma ...$
confidence \textit{regions}. When using $\chi _{1}^{2}$ the significance of $%
\Delta \chi ^{2}=32$ gets upgraded from a 5.4$\sigma $ to a 5.65$\sigma $
determination. More importantly, the reported errors on $s_{213}^{2}$ are
reduced from the 2-parameter $\Delta s_{213}^{2}=\pm 0.018$ to $\Delta
s_{213}^{2}=0.007$. The new errors, which are less than half the previous
ones, are equivalent to finding $\Delta s_{213}^{2}=1$ along the line $%
\epsilon =0$ intersecting the $\Delta \chi ^{2}=1$ contour of Figure \ref%
{fig:SimpleFit.eps}.

While it is possible to argue further, we do not find a one-parameter fit
convincing and we will \textit{not} choose to ignore $\epsilon$ to reduce
our error bars. It is not a question of setting $\Delta \chi^{2}=1$, or
plotting $\Delta \chi^{2}$ as a function of $s_{213}$. The issue is to make
the definition of the confidence region and the test being conducted
consistent. By starting with a hypothesis that extreme values of $\epsilon$
can invalidate the analysis we are committed to accounting for it as a
central parameter.

\subsubsection{Connection with the Daya Bay Analysis}

This is the first point where we notice an uncommon standard has entered the
neutrino literature. First, the texts of the $DB$ papers \cite{db1,db2} do
not specifically spell out the basis of their confidence level assignments.
No hypothesis is stated, yet two parameters of an absolute normalization $%
\epsilon$ and $\sin^{2}(2\theta_{13})$ are cited in their same paragraph.
How is $\Delta \chi^{2}$ being evaluated?

Figure 4 of Ref. \cite{db1}, reproduced as Fig. \ref{fig:DB_oscillation.eps}%
, allows one to deduce the method. The inset of the figure shows the $%
1\sigma, \, 3 \sigma, \, 5 \sigma$ variations occurring at $\Delta \chi^{2}
=1, \, 9, \, 25$. This established significance is being evaluated via $%
\chi_{\nu}^{2}$ with the number $\nu=1$.

If one were comparing the hypothesis of $s_{213}^{2} \neq 0$ (best fit) with 
$s_{213}^{2}=0$ (null) we would agree. The hypothesis of $s_{213}^{2} = 0$
has been ruled out. What is uncommon is to evaluate the uncertainty or
``standard error'' of $s_{213}^{2}$ also using $\chi_{\nu}^{2}$ with the
number $\nu=1$. There are 19 other background parameters floating to their
best fit values, developing a 20-dimensional error ellipsoid, while errors
have been found with no statistical penalties and using $\chi_{\nu}^{2}$
with the number $\nu=1$. Are all 19 parameters irrelevant nuisance
parameters, whose value has no bearing on the experiment?

$DB$ seems to suggests a penalty of a 2-parameter fit in citing $\chi_{Daya
\, Bay}^{2}/NF=4.6/4$, where $NF$ is the number of degrees of freedom, and
two parameters in the text. Besides citing $\sin^{2}(2 \theta_{13})$ they
write ``the absolute normalization $\varepsilon$ was determined from the fit
to the data.'' One might interpret $NF=4$ to mean 6 data points minus 2
"important" parameters ($\epsilon$ and $\sin^{2}(2\theta_{13})$). Actually
there are 24 terms in $\chi_{Daya \, Bay}^{2}$, which is fit with 20
parameters to give $NF=24-20=4$. The formula is:

\ba
\chi _{Daya\,Bay}^{2} &=&\sum_{d=1}^{6}\frac{\left[ M_{d}-T_{d}\left(
1+\varepsilon +\sum_{r}^{6}\,\omega _{r}^{d}\alpha _{r}+\varepsilon
_{d}\right) +\eta _{d}\right] ^{2}}{M_{d}+B_{d}}  \nonumber \\
&&+\sum_{r}^{6}\,\frac{\alpha _{r}^{2}}{\sigma _{r}^{2}}+\sum_{d=1}^{6}%
\left( \frac{\varepsilon _{d}^{2}}{\sigma _{d}^{2}}+\frac{\eta _{d}^{2}}{%
\sigma _{Bd}^{2}}\right) .  \label{chi2}
\ea

Symbol $T_{d}$ is the prediction from neutrino flux, $MC$ (simulations), and
neutrino oscillations, which involves integrating over the reactor energy
spectrum, and detector mass and acceptance, using a model.

The formula has 18 variationally-determined ``pull parameters'' in the set $%
\alpha_{r}, \, \varepsilon_{d}$, and $\eta_{d}$. The set of 20 parameters is
completed with $\varepsilon$ and $\sin^{2}(2 \theta_{13})$. Constants given
are $\sigma_{r}=0.8\%$, the uncorrelated reactor uncertainty, $%
\sigma_{d}=0.2\%$ the uncorrelated detection uncertainty. Symbol $B_{d}$ is
the background corresponding to data set $M_d$, and $\sigma_{Bd}$ the
background uncertainties of a few percent of the total number of events. We
added subscript $d$; the paper states that the values are given in a Table,
which unfortunately is not complete. The fraction of inverse beta decays
from the $r$-th reactor to the $d$-th detector as determined by baselines
and reactor fluxes is denoted $\omega_{r}^{d}$, a $6 \times 6$ array not
available from the paper or elsewhere.

Just as in our simple model fit, $DB$'s decision to use $\nu =1$ to assess a
20-parameter fit makes a difference in the definition of the confidence
level. The choice of $\nu =1$ has not been explained. Our analysis appears
to be the first to notice this decision might be questioned. For example, if
nobody cared about the value of $\epsilon $, $\alpha _{r},\,\varepsilon _{d}$%
, and $\eta _{d}$, and nobody looked to find them reasonable, we'd be very
surprised, yet agree with $\chi _{1}^{2}$ confidence levels, because we
don't have the authority to disagree. But for every parameter whose fit
value could have possibly invalidated the analysis, there is usually a
statistical penalty for introducing it, and reporting of fit values once
they are found. $DB$ evidently did examine 20 parameters with serious
concern for their values, writing \cite{db1} that "All best estimates of
pull parameters are within its (sic) 1 standard deviation based on
corresponding systematic uncertainties".

\section{The method of $\chi^{2}$ with pull}

\label{sec:chipull}

The method of \emph{$\chi^{2}$ minimization with pull parameters} ($%
\chi^{2}-pull$) was introduced for neutrino oscillation analysis about a
decade ago \cite{fogli1, sksolar, strumia, degouvea, huber03}. A related
reference is Stump \textit{et al.} \cite{stump}, Appendix B, which has been
cited by \cite{fogli1} and the $DB$ \cite{db1,db2}, RENO \cite{reno} and
Double Chooz \cite{dchooz2} papers. While there has been a breakthrough in
the technology of multiple detectors, this method of data analysis methods
does not specifically use it. The new results are also not solely
attributable to improved statistical errors. We noticed that the dramatic
improvement of precision claimed for $\theta_{13}$ measurements happened to
occur simultaneous with the use of the $\chi^{2}-pull$ analysis method.
Hence, the new claims cannot be directly compared to previous ones.

Recall the history. The CHOOZ null results of 1999 and 2003 \cite{chooz}
could not be surpassed for many years. Suddenly last year $DB$ reported $%
\sin^{2} 2 \theta_{13}=0.092 \pm 0.016_{stat} \pm 0.005_{syst},$ and
reported that $\sin^{2} 2 \theta_{13}=0$ was ruled out at the $5.2 \sigma$
confidence level mentioned above. The rapid advance in experimental
resolution came as a surprise to the community, even though the new Double
Chooz result \cite{dchooz1}, preceding $DB$ by a month or so, already showed
"indication for" disappearance of $\bar{\nu_{e}}$. Soon afterwards RENO \cite%
{reno} and Double Chooz \cite{ dchooz2,dchooz3} reported comparable
measurements with confidence levels ($CL$) of 4.9$\sigma$ and $2.9\sigma$
respectively. Almost overnight the reactor experiments had eclipsed the
expectations \cite{mn, g-chm, bmw, huber02} and results T2K \cite{t2k} and
MINOS \cite{ minosth13} of long baseline experiments , which had found only
indications of electron appearance consistent with 90\% $CL$. For reference,
a 90\% $CL$ translates to a 10\% chance a fluctuation in the null model
might give the value seen, while $5.2\sigma$ indicates a probability of $5.7
\times 10^{-7}$. In subsequent work Daya Bay \cite{db2} updated its
resolution to $7.7\sigma$, where the corresponding probability is $6 \times
10^{-13}.$

Above and beyond improvements in statistical errors, the sudden jump in
precision accompanied by a new data analysis method suggests that the method
itself is well worth exploring.

\subsection{Multi-Parameter Model Sensitivity}

The main characteristic of the $\chi^{2}-pull$ approach is the use of many
variationally-fit parameters and many additional terms not depending on data.

Daya Bay's fit uses 20 parameters applied to a sum of 24 squares involving 6
data points. RENO follows the same pattern, fitting 2 data points with 12
parameters. The number of parameters greatly exceeding the number of data
points does not seem to be widely appreciated. Careful reading (plus
checking for corroboration from members of the collaboration) is needed to
verify it is true \cite{private}. Meanwhile we find the experimental papers
do not provide sufficient detail to reproduce their calculations. Eighteen
of Daya Bay's fitting parameters are not reported.\footnote{%
Requests to DB for the full set of fitting parameters were denied.} We
decided to explore \textit{the analysis method} by making our own
calculations, as follows.

Consider a $\chi^{2}-pull$ model given by
\ba \chi_{eff}^{2} &=\sum_{d} \, {\frac{ (M_{d} +\eta_{d}- (1-s_{213}^{2}
s_{dE})M_{d 0}(1+\epsilon ) )^{2} }{M_{d}}} \nonumber \\
& + \sum_{d} \,{\frac{\eta_{d}^{2}}{(\sigma_{\eta} \sqrt{M_{d}})^{2}}} + {
\frac{(\epsilon-\epsilon_{0})^{2 } }{\sigma_{\epsilon}^{2}}} . \label{fat}
\ea The formula is a simplified version of $DB$'s statistic given in Eq. \ref
{chi2}. We note:

\begin{itemize}
\item The formula uses 6 parameters $\eta_{1}, \, \eta_{2}...\eta_{6}$ to
emulate those seen in the literature. Each parameter $\eta_{j}$ is in
principle capable of tuning $\chi^{2}$ arbitrarily close to zero for the
corresponding term. We found this feature to be crucial for explaining how
the $\chi^{2}-pull$ method works. The balancing ``force'' that prevents a
trivial fit and $\chi^{2}=0$ is produced by the added terms going like $%
\eta_{d}^{2}/\sigma_{\eta}^{2}$, which do not depend on the data. The tug-of
war between the two terms is regulated by ``pull denominators'' such as $%
\sigma_{\eta}^{2}.$ These denominators then develop a crucial role in the
outcomes.

\item We have scaled the pull denominators regulating $\eta_{j}$ to be $%
(\sigma_{\eta}\sqrt{M_{d}})^{2}$. Assuming a typical statistical fluctuation
of order $\sqrt{M_{d}}$ detector-by-detector puts a natural scale of $%
\sigma_{\eta}$ at order unity. One parameter $\sigma_{\eta}$ then suffices
to parameterize backgrounds that scale at the same order as statistical
noise, detector by detector.

\item There are two interpretations of $\epsilon_{0}$. By a shift of
variables $\epsilon \rightarrow \epsilon-\epsilon_{0}$ the $\epsilon_{0}$
term appears inside the expression involving $M_{d 0}$. Then $\epsilon_{0}$
can stand for the accumulated constants of all the other parameters.

\item The more physical interpretation goes back to $\chi^{2}-pull$ as a
maximum log-likelihood estimator, discussed in Ref. \cite{stump}. Adding to $%
\chi^{2}$ a term going like ${\epsilon^{2 }/\sigma_{\epsilon}^{2}}$
corresponds to multiplying a distribution by a ``prior distribution'' factor
of $exp({-\epsilon ^{2 } /2\sigma_{\epsilon}^{2}})$. It is interesting that
all the priors of the $\chi^{2}-pull$ methods in use happen to be centered
on zero. For example (recall Eq. \ref{chi2}) priors going like $%
exp(-\eta_{d}^{2}/\sigma_{Bd}^{2})$ make a model in which half the
probability describes \textit{negative backgrounds}. Perhaps this might be
improved \cite{fc}. For our purposes the act of bundling together the
cumulative effects of 12 parameters into one parameter $\epsilon_{0}$
suffices to produce an illustrative model.
\end{itemize}

Once again we state our hypothesis. We propose to test the null model $%
s_{213}=0$, and all other parameters are of order $1\% \pm 1\%$. Being more
specific, with a Bayesian prior distribution of these parameters, is
certainly possible but a side issue. We intend to pay a statistical penalty
for using extra parameters that might invalidate our test.

\subsection{Analysis Results and Analysis Sensitivity}

Table \ref{tab:fitvals} shows several results of minimizing Eq. \ref{fat} to
fit our data set while exploring a range of parameter values. We notice that
a wide range of different values of $\chi ^{2}$ and $s_{213}^{2}$ are
possible from one data set using the $\chi ^{2}-pull$ method. That is, the
method is highly sensitive to small perturbations of model parameters. It is
exactly what Stump \textit{et al.} warned with \textquotedblleft small
inaccuracies in the (systematic error) values...may translate into a large
error on the confidence levels computed from the chi-squared distribution 
\cite{stump}.\textquotedblright\ What causes the great sensitivity to free
parameters? It turns out that the denominators of the pull-terms control a
great deal.

\begin{table*}[htbp]
\begin{center}
\begin{tabular}{ccccccccccc}
$\eta_{1}$ & $\eta_{2} $ & $\eta_{3} $ & $\eta_{4} $ & $\eta_{5} $ & $%
\eta_{6} $ & $\epsilon $ & $\epsilon_{0}$ & $\sigma_{\epsilon}$ & $%
sin^{2}(2\theta_{13})$ & $\chi_{eff}^{2}$ \\ \hline
1.87549 & -2.20624 & -1.79852 & -0.221603 & 0.182851 & 0. 746333 & 
-0.00658901 & -0.01 & 0.005 & 0.06 & 6.92403 \\ 
2.05348 & -2.05461 & -2.01523 & -0.784737 & -0.390732 & 0 .186919 & 
-0.00265046 & -0.01 & 0.005 & 0.08 & 9.44853 \\ 
2.21698 & -1.92084 & -2.25033 & -1.35392 & -0.96955 & -0. 38150 & 0.00126211
& -0.01 & 0.005 & 0.1 & 19.2687 \\ 
2.74927 & -1.32105 & -1.12067 & -0.118705 & 0.28375 & 0.8 51055 & -0.00348295
& 0. & 0.005 & 0.06 & 6.95282 \\ 
2.92984 & -1.16492 & -1.33641 & -0.682966 & -0.288892 & 0 .288097 & 
0.000480337 & 0. & 0.005 & 0.08 & 6.31658 \\ 
3.09575 & -1.02374 & -1.56628 & -1.25340 & -0.867737 & -0. 281666 & 
0.00441777 & 0. & 0.005 & 0.1 & 12.9967 \\ 
3.62164 & -0.435221 & -0.443855 & -0.0154563 & 0.387047 & 0.953269 & 
-0.000376915 & 0.01 & 0.005 & 0.06 & 12.4968 \\ 
3.80611 & -0.275163 & -0.657549 & -0.581041 & -0.186864 & 0.388763 & 
0.00361113 & 0.01 & 0.005 & 0.08 & 8.67999 \\ 
3.97628 & -0.126413 & -0.882846 & -1.15238 & -0.766739 & -0.181635 & 
0.00757351 & 0.01 & 0.005 & 0.1 & 12.2003 \\ 
0 & 0 & 0 & 0 & 0 & 0 & -0.0022265 & 0 & N/A & 0.069828 & 5.6563 \\ 
0 & 0 & 0 & 0 & 0 & 0 & -0.022263 & 0 & N/A & 0 & 37.605%
\end{tabular}%
\end{center}
\caption{Parameters of some fits shown in Figure \ref%
{fig:Sig005MultiSig1Multi.eps}, listed so they can be reproduced. The last
two rows are the straightforward simple model.}
\label{tab:fitvals}
\end{table*}

\begin{figure}[htbp]
\begin{center}
\includegraphics[width=3in]{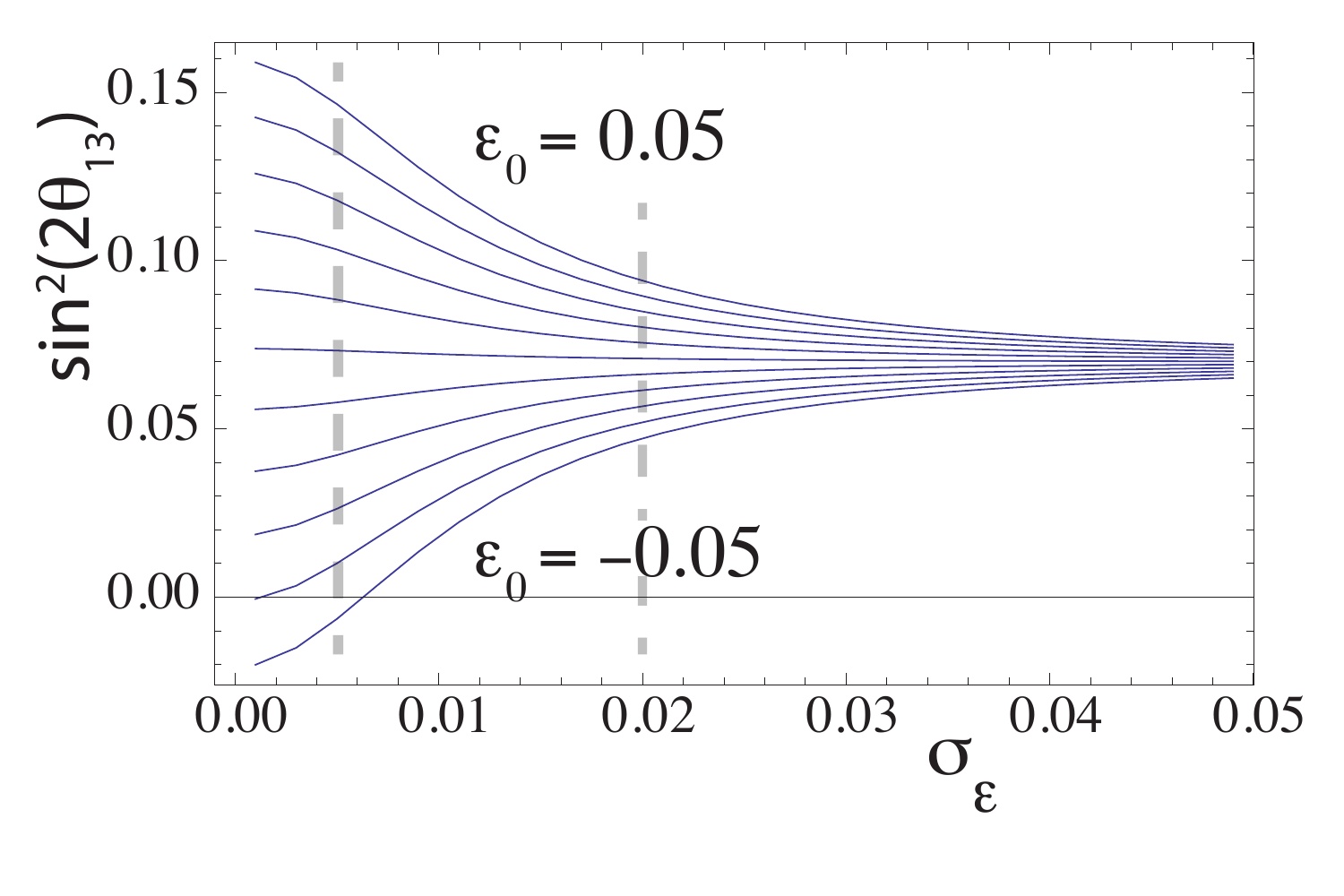}
\end{center}
\caption{ The best fit value of $\sin^{2}(2 \theta_{13})$ as a
function of the pull-parameter uncertainty $\sigma_{\epsilon} $ for different $\epsilon_{0}$
 ranging
from -0.05 to 0.05 in steps of 0.01. The transparent dashed vertical lines
cutting across the plot indicates the values used to make the two panels of
Figure \ref{fig:Sig005MultiSig1Multi.eps} }
\label{fig:SSqVSig0505.eps}
\end{figure}

Fig \ref{fig:SSqVSig0505.eps} shows the dependence of the central value of $%
s_{213}^2$ on the pull-parameter uncertainty $\sigma_{\epsilon}$. The
different curves use $\epsilon_{0}$ ranging from -0.05 to 0.05 in steps of
0.01. The other pull parameter uncertainty $\sigma_{\eta} $ is fixed at 0.1.
Note the significant variation and (we believe) unacceptable sensitivity as $%
\sigma_{\epsilon} \rightarrow 0.$ For sufficiently large $\sigma_{\epsilon}$
the sensitivity of the central value of $s_{213}^2$ actually disappears. The
reasons are trivial from inspecting the formula. It is rather important that
the error $\Delta s_{213}^2$ increases in the same limit.

\begin{figure}[tbp]
\begin{center}
\includegraphics[width=3in]{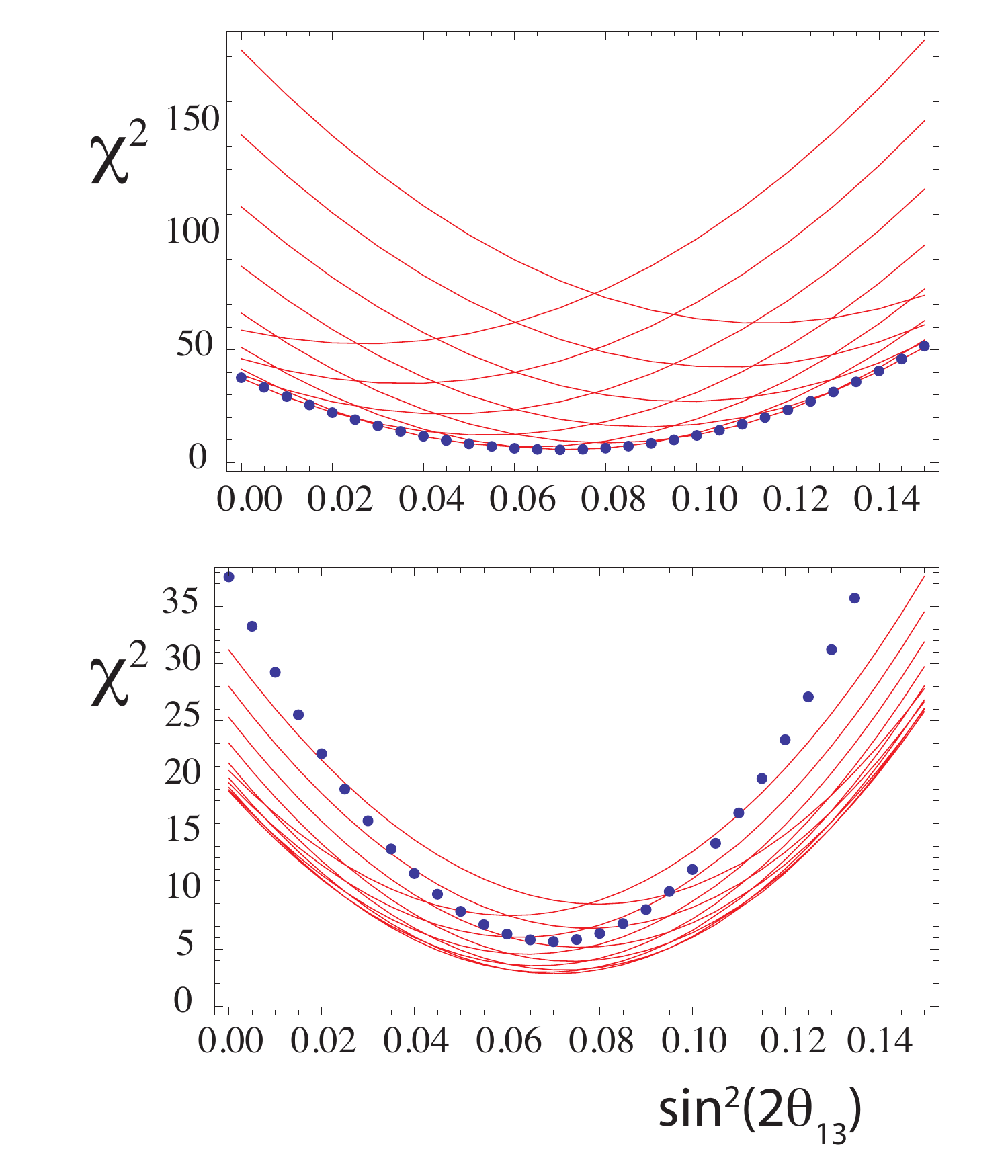}
\end{center}
\caption{High sensitivity of the $\chi^{2}-pull$ method to small
parameters. Both panels show $\chi_{eff}^{2}$ as a function of $%
s_{213}^2$. Upper panel: $\sigma_{\eta}=0.1$ and $%
\sigma_{\epsilon} =0.005$. Lower panel: $\sigma_{\eta%
}=1$ and $\sigma_{\epsilon}=0.02$. The dots are $\chi%
^{2}$ of the simple model, Eq. \ref{simple}. }
\label{fig:Sig005MultiSig1Multi.eps}
\end{figure}

Figure \ref{fig:Sig005MultiSig1Multi.eps} shows $\chi _{eff}^{2}$ as a
function of $s_{213}^{2}$ with all other parameters floating to their
minimizing values. As above the different curves are $\epsilon _{0}$ ranging
from -0.05 to 0.05 in steps of 0.01. The upper panel plot uses $\sigma
_{\eta }=0.1$ and $\sigma _{\epsilon }=0.005$, the region of large
dispersion in Figure \ref{fig:SSqVSig0505.eps}. Such a small value of $%
\sigma _{\epsilon }$ tends to prevent the parameters from improving $\chi
^{2}$. That is shown by the dots along the bottom of the plot, which
represent the fit to the simple model, Eq. \ref{simple}. Most of the
envelope of $\chi ^{2}$ values tend to be bounded above the simple model. We
find this is significant.

The lower panel of Figure \ref{fig:Sig005MultiSig1Multi.eps} shows the same
plot when $\sigma_{\eta}=1$ and $\sigma_{\epsilon}=0.02$. Those choices
allows greater freedom for $\eta_{d}$ to improve the fit. Actually the
improvement in $\chi^{2}$ from this parameter is marginal. However there is
a dramatic change in the \textit{width} of the plot (and the precision $%
\Delta s_{213}^2$) compared to the other case. Despite the importance of $%
\sin^{2}(2 \theta_{13})$ and its uncertainty, we have seen no specific
discussion of the wide range of results that can be obtained simply by
adjusting the pull-width parameters. Our hypothesis to fit data with the
method must be abandoned because the range of possible values we can find
greatly exceeds the range of any error bars from the same analysis.

%
%

\subsection{Sensitivity Explained: Built-In Degeneracies}

Once the pull-width parameters are set somehow, attention shifts to the
change in $\chi^{2}$ from varying pull parameters near the minimum. A lack
of sensitivity of $\chi^{2}$ has been promoted as a virtue. Actually it is a
symptom of analysis degrading resolution.

Consider a general function $\chi^{2}(\theta)$ depending on parameters $%
\theta =\theta_{1}...\theta_{n}$. Find the best fit points with $\partial
\chi^{2}/\partial \theta_{i}=0$ at $\theta=\theta_{*}$. The curvature at $%
\theta_{*}$ is $\a_{ij}={\frac{1}{2}}(\partial \chi^{2}/\partial
\theta_{i}\partial \theta_{j})_{\theta_*}$. The value of $\chi^{2}$ and its
curvature depend on the number of terms and parameters. For example, RENO's
formula \cite{reno} for $\chi^{2}$ has 12 terms and 12 parameters, possibly
explaining why the minimum $\chi^{2}$ shown in the paper's Figure 3 is 
\textit{zero}. When there are even more parameters than terms to be fit,
then $\chi^{2}$ must be \textit{totally insensitive} to certain linear
combinations of the parameters. Insensitivity means the curvature
eigenvalues will be unusually small from built-in near degeneracies. But it
is not strictly necessary to have more parameters than terms. It is
sufficient for the nature of the parameters to nearly reproduce one another
to produce built-in near degeneracies.

The \textit{covariance} matrix $C_{ij}=(\a^{-1})_{ij}$ has the inverse
eigenvalues. It defines the standard uncertainty of fit parameters. For
example the uncertainty of $\theta _{1}$ when $\chi ^{2}$ varies by $\Delta
\chi ^{2}$ in a one-parameter variation is $\sqrt{C_{11}\Delta \chi ^{2}}$.
Inverting matrices with a large ratio of maximum eigenvalues ($\lambda
_{max} $) to minimum eigenvalues ($\lambda _{min}$) is unstable, also called
\textquotedblleft ill-conditioned\textquotedblright . An ill-conditioned
inverse has inordinate sensitivity to small projections of parameters along
the eigenvectors with small eigenvalues. The cure for ill-conditioned
problems removes the subspace of $a_{ij}$ labeled by eigenvalues below some $%
cut$ value by inverting the matrix in the complementary subspace, forming
the \textit{pseudoinverse} $C_{cut}$. In symbols \ba \a & =\sum_{\a}^{K} \, |a_{\a}> \lambda_{\a} <a_{\a}|; \nn \\ C_{cut}  & = \sum_{\a}  \, \theta( \lambda_{\a}/\lambda_{max} >cut)  |a_{\a}> (\lambda_{\a})^{-1} <a_{\a}|.  \nn \ea The step function $\theta( \lambda_{\a}/\lambda_{max} >cut)$ enforces the eigenvalues $\lambda_{\a}$ are larger than the fraction $cut$ of the largest eigenvalue $\lambda_{max}$. 

\begin{figure}[tbp]
\begin{center}
\includegraphics[width=3in]{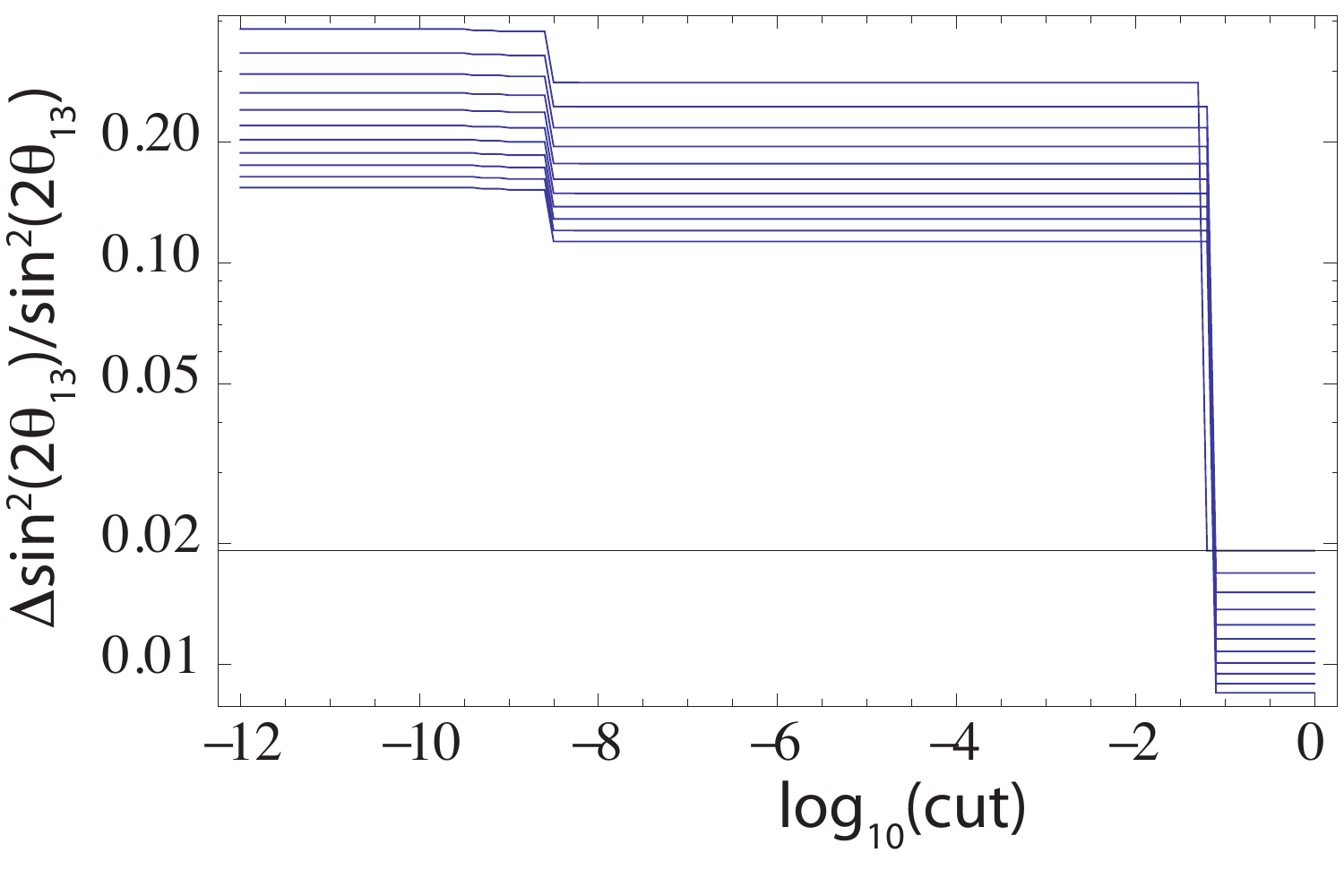}
\end{center}
\caption{A case of high sensitivity of the $\chi^{2}-pull$ method
explained. The figure shows the relative uncertainty $\Delta
s_{213}^2/s_{213}^2$ evaluated as a function of the $logcut =\log_{10}(%
\lambda_{min}/\lambda_{max})$, where $\lambda_{min}$
and $\lambda_{max}$ are the minimum and maximum eigenvalues retained
in making the covariance matrix. Different curves represent $\epsilon%
_{0}$ ranging from -0.05 to 0.05 in steps of 0.01. }
\label{fig:Eta01Sig021IllCuts}
\end{figure}

Figure \ref{fig:Eta01Sig021IllCuts} shows how the relative uncertainty $%
\Delta s_{213}^{2}/s_{213}^{2}$ depends on the ratio $logcut=-\log
_{10}(\lambda _{max}/\lambda _{min})$. (For simplicity $\Delta \chi ^{2}=1$;
the calculation can be trivially rescaled for a more conventional value.)
Note the logarithmic scale. The calculation uses $\sigma _{\eta }=1$, $%
\sigma _{\epsilon }=0.01$, and explores the range $-0.05\leq \epsilon
_{0}\leq 0.05$ in steps of 0.01. The steps in $\Delta
s_{213}^{2}/s_{213}^{2} $ come at the $logcut$ values where an eigenspace
and its corresponding contribution to the calculation is removed. Very tiny
variations produce sudden and large effects, the classical signal of an
ill-conditioned problem. Many cases of $\chi ^{2}-pull$ that we investigated
were ill-conditioned.

That explains the ``paradox'' cited in the introduction: the high
sensitivity of fitted results to external parameters is found precisely in
the \textit{insensitivity} of $\chi^{2}$ values to an excessive number of
variationally-fit parameters. Our investigation has found that using too
many parameters creates an ill-conditioned, insensitive procedure that
actually decreases the resolution of $\sin^{2}(2 \theta_{13})$. A simpler
analysis targeted on the physical parameter $s_{213}^{2}$ \textit{must
necessarily} improve the precision of its determination. For illustration of
this, compare the simple straightforward model of Section \ref{sec:straight}.

\section{Summary}

\label{sec:discuss} We began this study as a sort of detective investigation
to discover what had been done with neutrino reactor data. Our initial
impression was that straightforward data analysis took maximum advantage of
the new technology of identical detectors. Actually we found that the
analysis methods in wide use, by becoming so complicated, have not come
close to optimizing the precision of $s_{213}^{2}$.

What is gained by a multi-parameter fit? Compared to our simple fit, the
accomplishment from 18 additional parameters used by $DB$ is a change of $%
\chi^{2}$ by 1.3 units. Suppose one finds a logical argument (which we've
not seen) for classifying all parameters as irrelevant nuisances, except the
mixing angle of great interest. Then using $\chi_{1}^{2}$ to evaluate $%
\Delta \chi^{2}$ would be appropriate. Meanwhile the hidden penalty of 18
extra parameters tends to decrease the precision of $s_{213}^2$, not
increase it. We don't find that necessary or welcome.

Multi-parameter studies are common in simulating and de-bugging hardware. We
believe the information from Daya Bay's 20 parameter fit is that numerous
parameters that might have detected significant systematic errors were found
to be small. This is a guess about parameter information those in the field
need to know. (The facts are actually unknown so long as the full set of
fitting parameters is unavailable.) If the guess regarding systematic errors
is correct, it is a wonderful result indicating brilliantly constructed
hardware. Physicists interested in neutrino physics know and expect that
internal studies of hardware systematics might involve 10 or 20 or 100
parameters laboriously checked and double-checked. And this is supposed to
be done before ``opening the box'' to look at the physical parameter of
interest. There is no logical necessity to mix the two different goals.
Indeed it would be disappointing if the level of analysis of systematic
errors by $DB$ and RENO consisted solely of the unconvincing method (in our
opinion) of $\chi^{2}-pull$ presented in published work.

To reiterate this, our analysis finds that a hardware study is far from the
best way to fit $s_{213}^2$. Once systematic errors \textit{are known
independently and with good reliability} to be small, a few-parameter method
tends to be a more effective way to evaluate $s_{213}^2$ and its errors.
This, and the previous material, explain the remark earlier that the errors
in $s_{213}^2$ can be viewed as both under-estimated and over-estimated. If
outsiders to the experiment were to present a 20-parameter fit, we believe
no credence would be given unless the results were assessed with $%
\chi_{20}^{2}$, if not an even more demanding standard. The significance of
detecting $s_{213}^2 \neq 0$ with 32 units of $\Delta \chi^{2}$ would then
be reduced to an equivalent $2 \sigma$ effect, and errors computed using $%
\Delta \chi^{2}=1$ would be considered \textit{under}-estimated. Meanwhile
using 20 parameters has also so flattened the $\chi^{2}$ function by near
degeneracies, that it has diluted the impact of data on the measurement. On
that basis it is an approach wastefully \textit{over}-estimating the errors
of the competing physical parameter.

We found $s_{213}^2= 0.07 \pm 0.018$ with $\chi^{2}= 5.66$ at $\epsilon =
-0.002$. On its face this is a better fit than $DB$'s. Our result is
illustrative and certainly not the last word, but it strongly suggests that
even better methods must exist. It would be good for the experimental groups
to present straightforward, simple fits where all definitions are complete,
all variables are reported, data is divulged, and results are reproducible.
Simplicity and transparency will greatly assist the main interest in
neutrino data, which is the comparison of experiments with competing models
of the underlying physics. \newline

\textbf{Acknowledgements:} We thank Daya Bay collaboration members J. Cao,
K. Heeger, P. Huber, W. Wang and K. Whisnant for information about the
experiment and the analysis procedures. We also thank Danny Marfatia for
helpful information. Research supported in part under DOE Grant number
DE-FG02-04ER14308. ANK thanks Professor F. Tahir for consultation and
advice, the University of Kansas Theory Group for kind hospitality, the
Higher Education Commission of Pakistan for support for graduate studies and
travel under the Indigenous Ph.D. Fellowship Program Batch-IV and
International Research Support Initiative Program.

\section{Appendix: Extracting Data}

\begin{figure}[htbp]
\begin{center}
\includegraphics[width=3in,height=2in]{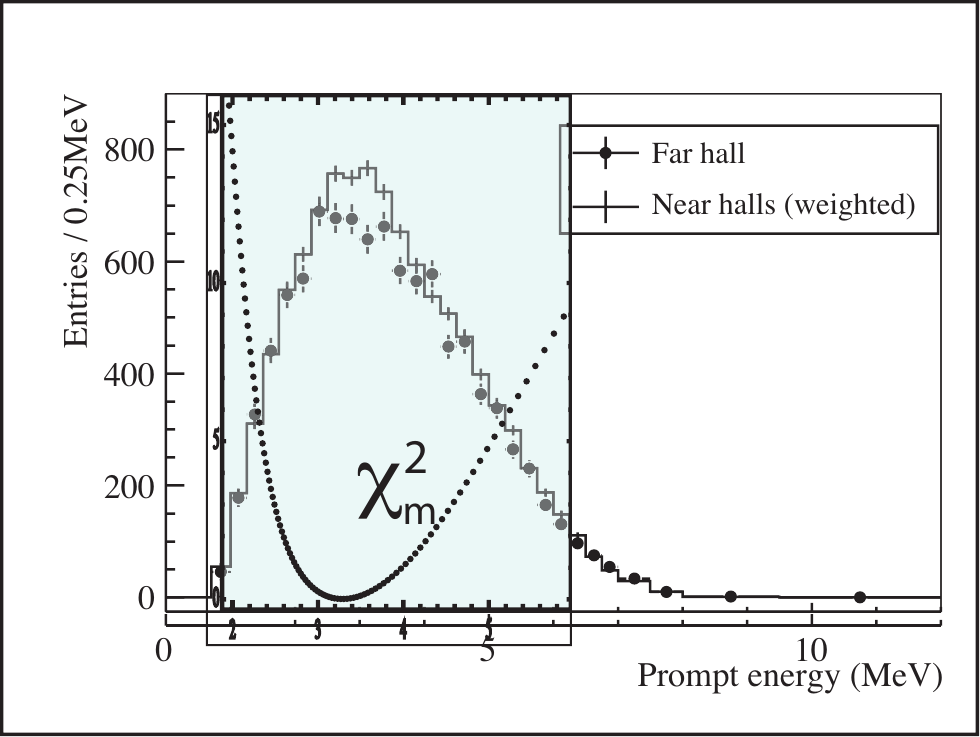}
\end{center}
\caption{ {\small Goodness of fit $\chi_{m}^2$ between a
2-parameter oscillation model and the curve of Figure \ref%
{fig:DB_oscillation.eps} is shown as dots. The horizontal axis is the
neutrino energy $E$ in MeV. The histogram and points with error bars are the
neutrino energy spectrum from Figure 5 of Ref \cite{db1}, superposed
with its scale. The minimum of $\chi^2_m $ at the peak of the energy
spectrum is a consistency check that also effectively finds $\Delta
m_{13}^{2}=2.32 \times 10^{-3} eV^2$. }}
\label{fig:DYB_spectrumOverlaps}
\end{figure}

The definitions of quantities and their values accompanying Eq. \ref{chi2}
are taken directly from the experimental report \cite{db1,db2}. The reports
may appear to define everything, but the information is incomplete. We use
the following strategy to fill in gaps. From its usage $T_{d} \propto
<1-\sin^{2}2\theta_{13}<\sin^{2}(\Delta m_{31}^{2}L_{d}/E)>/L_{d}^{2}$,
where the reactor-detector separation is $L_{d}$ and the angle-brackets
represent an average over the reactor spectrum multiplied by the cross
section and acceptance \footnote{%
We ignore a negligible correction involving $\theta_{12}$}. Those details
might seem to preclude any challenge. Fortunately the effective survival
probability $P_{eff}$ for $DB$ is given in the paper's Figure 4, reproduced
here as Fig. \ref{fig:DB_oscillation.eps}. The red curve shows the fraction $%
P_{d}$ of neutrinos surviving at best fit. It is important that \textit{the
curve already includes all reactor and detector effects: flux, distance,
acceptance, live time, efficiency, backgrounds, etc. }

Using $P_{d}$ and the expected events listed in $DB$'s Table II produced our
Table I. There are other ways to proceed. $DB$ lists inverse beta decay
candidate event rates per day, live times, and some background figures.
Combining those number produces numbers close to those of Table I, but not
exactly the same. Since we are making points of principle and procedure the
exact numbers are important but not crucial. For one thing, $DB$'s
description of efficiencies and ``simulations'' are not described in enough
detail to reproduce their values.

Digitizing the $P_{d}$ curve in Fig. \ref{fig:DB_oscillation.eps} and
fitting it produces 
\[
P_{eff}(L_{eff},\theta ,\mu )=1-\sin ^{2}(2\theta )\sin ^{2}(\mu
L_{eff}/km), 
\]%
where $\sin ^{2}(2\theta )=0.0762$, $\mu =0.894$. Here $L_{eff}$ is the
effective distance defined in computing the curve. For the entire range
covering the data the differences of the 2-parameter fit from the curve are
less than the thickness of the curve. (Actually, we only need the $L_{eff}$
dependence and $P_{d}$ evaluated at the detector positions. The good quality
of the curve fit is a consistency check.)

As a further check, recall that mono-energetic neutrinos of energy $E$
mixing by one angle predict $\mu =1.267(\Delta m_{13}^{2}/eV^{2})E/GeV$.
Using $\Delta m_{13}^{2}=2.32\times 10^{-3}eV^{2}$ translates the
2-parameter fit into an acceptance-averaged energy of $E=3.3$ MeV. It is
quite significant that this average energy precisely matches the peak of the
spectrum, as shown in Figure \ref{fig:DYB_spectrumOverlaps}. To make the
figure we defined $\chi _{m}^{2}(E)$ by summing the differences of squares
of the curve and fit over 141 digitized values, and finally multiplying by
an arbitrary factor of 100 to make the values visible. This approach to
fitting $\mu $ also finds $\Delta m_{13}^{2}$ independently using the $DB$
data \footnote{%
While finding $\Delta m_{13}^{2}$ is important, the analysis is dependent on
the $DB$ figure we digitized, which uses a model described in the paper.}.
The data shown by the probabilities in Fig. \ref{fig:DB_oscillation.eps}
detector-by-detector is now ready to be fit.


\begin{thebibliography}{99}
\bibitem{private} In the course of this research, we checked over a period
of several months with helpful Daya Bay Collaboration members, and it was
confirmed that 20 parameters are indeed used and the error $\Delta \sin
^{2}(2\theta _{13})$ is evaluated with parameters floating to their best-fit
values.
\end{thebibliography}
\end{document}